\documentclass[3p,times,procedia]{elsarticle}
\usepackage{nupha_ecrc}

\volume{00}

\firstpage{1}

\journalname{Nuclear Physics A}

\runauth{S. R. Klein}

\jid{nupha}

\jnltitlelogo{Nuclear Physics A}

\usepackage{amssymb}

\usepackage[figuresright]{rotating}

\begin{document}
\begin{frontmatter}



\dochead{}

\title{Ultra-peripheral collisions and hadronic structure}


\author{Spencer R. Klein}

\address{50R5008 Lawrence Berkeley National Laboratory, 1 Cyclotron Road, Berkeley CA 94720 USA}

\begin{abstract}
Ultra-peripheral collisions are the energy frontier for photon-mediated interactions, reaching, at the Large Hadron Collider (LHC),  $\gamma-p$ center of mass energies five to ten times higher than at HERA and reaching $\gamma\gamma$ energies higher than at LEP.    Photoproduction  of  heavy quarkonium and dijets in $pp$ and $pA$ collisions probes the gluon distribution  in protons at Bjorken-$x$ values down to $3\times10^{-6}$, far smaller than can be otherwise studied.  In $AA$ collisions, these reactions probe the gluon distributions in heavy ions, down to $x$ values of a few $10^{-5}$.    Although more theoretical work is needed to nail down all of the uncertainties, inclusion of these data in current parton distribution function fits would greatly improve the accuracy of the gluon distributions at low Bjorken-$x$ and low/moderate $Q^2$. 
High-statistics $\rho^0$ data probe the spatial distribution of the interaction sites; the site distribution is given by the Fourier transform of $d\sigma/dt$.  
After introducing UPCs, this review presents recent measurements of dilepton production and light-by-light scattering and recent data on proton and heavy nuclei structure, emphasizing results presented at Quark Matter 2017 (QM2017). 
\end{abstract}


\end{frontmatter}


\section{Introduction}
\label{Sec:int}

Relativistic heavy ions are accompanied by strong electromagnetic fields.  The electric and magnetic fields are perpendicular, so in the Weizs\"acker-Williams approach, the combined fields may be treated as a flux of nearly-real virtual photons.  The maximum photon energy depends on the width of the fields in the direction of motion: $k_{\rm max}=2\gamma\hbar c/b$, where $\gamma$ is the Lorentz boost of the ion, and b is the transverse distance from the ion (impact parameter).

UPCs are interactions between the electromagnetic fields of one ion and the other ion or its electromagnetic field \cite{Bertulani:1987tz,Bertulani:2005ru,Baltz:2007kq}.  Usually, UPCs are studied as exclusive reactions, unaccompanied by particles from hadronic interactions.      However, ALICE and STAR  have recently presented data on $e^+e^-$ pairs in peripheral hadronic collisions, exhibiting the characteristic signature of coherent electromagnetic production: a large peak at low transverse momentum ($p_T$). 

UPCs provide the highest energy electromagnetic probes available at particle accelerators.   At the LHC, UPCs reach $\gamma p$ energies up to 2 TeV - ten times higher than were available at HERA, and sufficient to probe gluon distributions at Bjorken-$x$ values down to $3\times10^{-6}$.  For heavy-ion targets, the only previous data with electromagnetic probes is from fixed-target experiments;. The new UPC data is directly sensitive to nuclear shadowing of gluon distributions down to Bjorken$-x$ values down to about $10^{-5}$.


\section{Dileptons}
\label{Sec:dil}

Purely electromagnetic dilepton production is important, both as a calibration tool and for physics studies.  ALICE, ATLAS and STAR have all reported high-statistics samples.  Figure 1 shows the rapidity distribution of the ATLAS dimuon sample \cite{ATLASmumu} from proton-lead collisions along with the mass distribution for lead-lead collisions, compared to the predictions of the STARlight Monte Carlo \cite{Klein:2016yzr}.  The agreement is excellent, especially considering that STARlight uses the equivalent photon approximation with a lowest-order calculation \cite{Baltz:2009jk}.   Higher order corrections are not important for these dimuons.

At QM2017, ATLAS presented a particularly interesting process: light-by-light scattering, $\gamma\gamma\rightarrow\gamma\gamma$  \cite{ATLASmumu, Aaboud:2017bwk}. It  occurs only via a higher-order quantum process, described by a box diagram (charge particle loop) connecting the incident and outgoing photons.  Because all  charged particles contribute to the loop, the process is sensitive to new, beyond-standard-model charged particles.  ATLAS found 13 events within their kinematic selection and with pair mass above 6 GeV, compared to an estimated background of $2.6\pm0.7$ events.  The resulting cross-section, 70 $\pm\ 24$ (stat.) $\pm\ 17$ (syst.) nb, is compatible with the standard model prediction \cite{dEnterria:2013zqi}.  The current statistical precision is not yet sensitive to beyond-standard-model physics.

\begin{figure}[htb]
\center{\includegraphics[width=0.36\columnwidth]{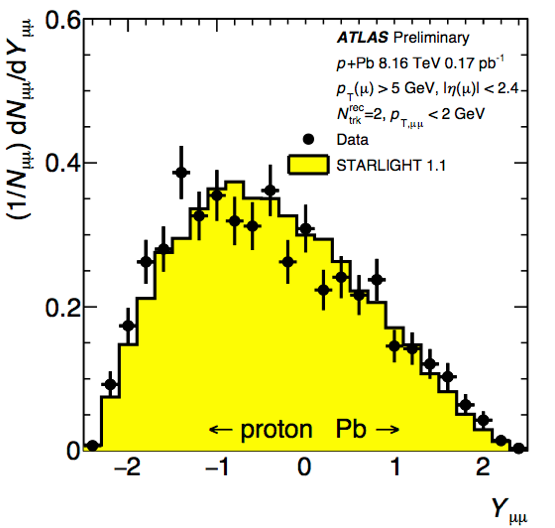} 
\includegraphics[width=0.4\columnwidth]{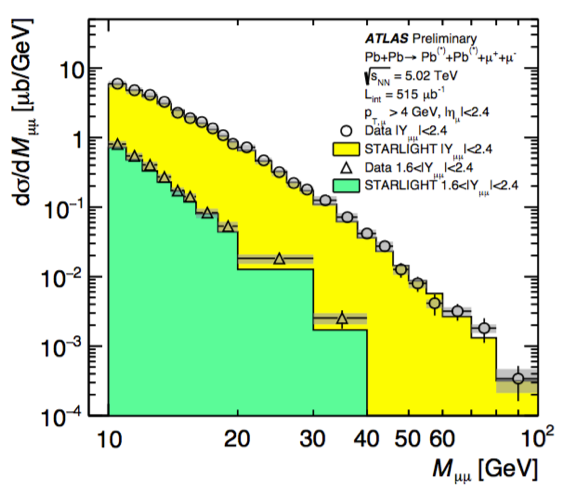} }
\label{fig:pairs}
\caption{(left) The rapidity distribution for exclusive $\mu^+\mu^-$ pairs observed in proton-lead collisions seen in ATLAS at $\sqrt{s_{NN}} = 8.16$ TeV.  (right) The mass distribution for exclusive $\mu^+\mu^-$ pairs observed by the ATLAS collaboration in lead-lead collisions.  Both distributions are in excellent agreement with STARlight \cite{Klein:2016yzr} simulations. From Ref. \cite{ATLASmumu}.  
}
\end{figure}

UPCs also have implications for accelerator design.  In Bound-Free Pair Production (BFPP), the electron is produced bound to one of the incident nuclei \cite{Bertulani:1987tz}.   For lead ions at the LHC, the cross-section for BFPP is about 200 b \cite{Meier:2000ga}, so it contributes significantly to the loss of ions from the beam.   The ion momenta are nearly unchanged, but their charge is reduced by one, so the single-electron ions gradually separate from the main beam   \cite{Klein:2000ba}.  At the LHC,  lead ions strike the beampipe about 135 m downstream from the interaction points, depositing their energy in a hadronic shower.  In a recent test at a luminosity of $2.3\times 10^{27}$/cm$^2$/s, the beam deposited about 53 Watts of power, enough energy to quench the dipole magnet that it struck \cite{Jowett:2016yfa}.   The impact of these BFPP beams can be mitigated by using  orbit bumps, but, even with this BFPP will nevertheless impact high luminosity operations and limit future, higher energy colliders. 

\section{Vector meson photoproduction}

Vector meson photoproduction is the most intensively studied UPC reaction.   The cross-sections are large, most of the final states are simple, and the process holds considerable theoretical interest.   Cross-sections have been measured for  $\rho$, $\omega$, one or more $\rho'$ states,  $J/\psi$ and $\psi'$, and $\Upsilon$ states, along with directly produced $\pi^+\pi^-$.   The photon energy $k$ is related to the final state mass $M_V$ and rapidity $y$:
\begin{equation}
k = \frac{M_V}{2}\exp{(\pm y)}
\end{equation}
The $\pm$ sign reflects the two-fold ambiguity as to which nucleus emits the photon and which is the target.  

In photoproduction, the incident photon fluctuates to a quark-antiquark pair which then scatters quasi-elastically from a target nucleon, emerging as a vector meson.  The scattering is mediated by the strong force, but occurs without a net color exchange.  At low photon energies, this proceeds by meson exchange, with a cross-section that decreases with increasing $k$.  At higher energies, photoproduction occurs primarily via Pomeron exchange, with a cross-section that rises slowly with increasing $k$.   The Pomeron represents the absorptive part of the cross-section.   In QCD, it is a gluon ladder. To lowest order, it consists of two gluons.

Dipions ($\pi^+\pi^-$) can be photoproduced directly, or through the decays of the $\rho$ and $\omega$.  In direct production, the photon fluctuates directly to a pion pair.  Since these channels are indistinguishable, they interfere with each other.  STAR and ALICE have made high-statistics measurements of dipion photoproduction.  In STAR, the trigger required that the $\rho^0$ be accompanied by mutual Coulomb excitation. 

\begin{figure}[htb]
\center{\includegraphics[width=0.47\columnwidth]{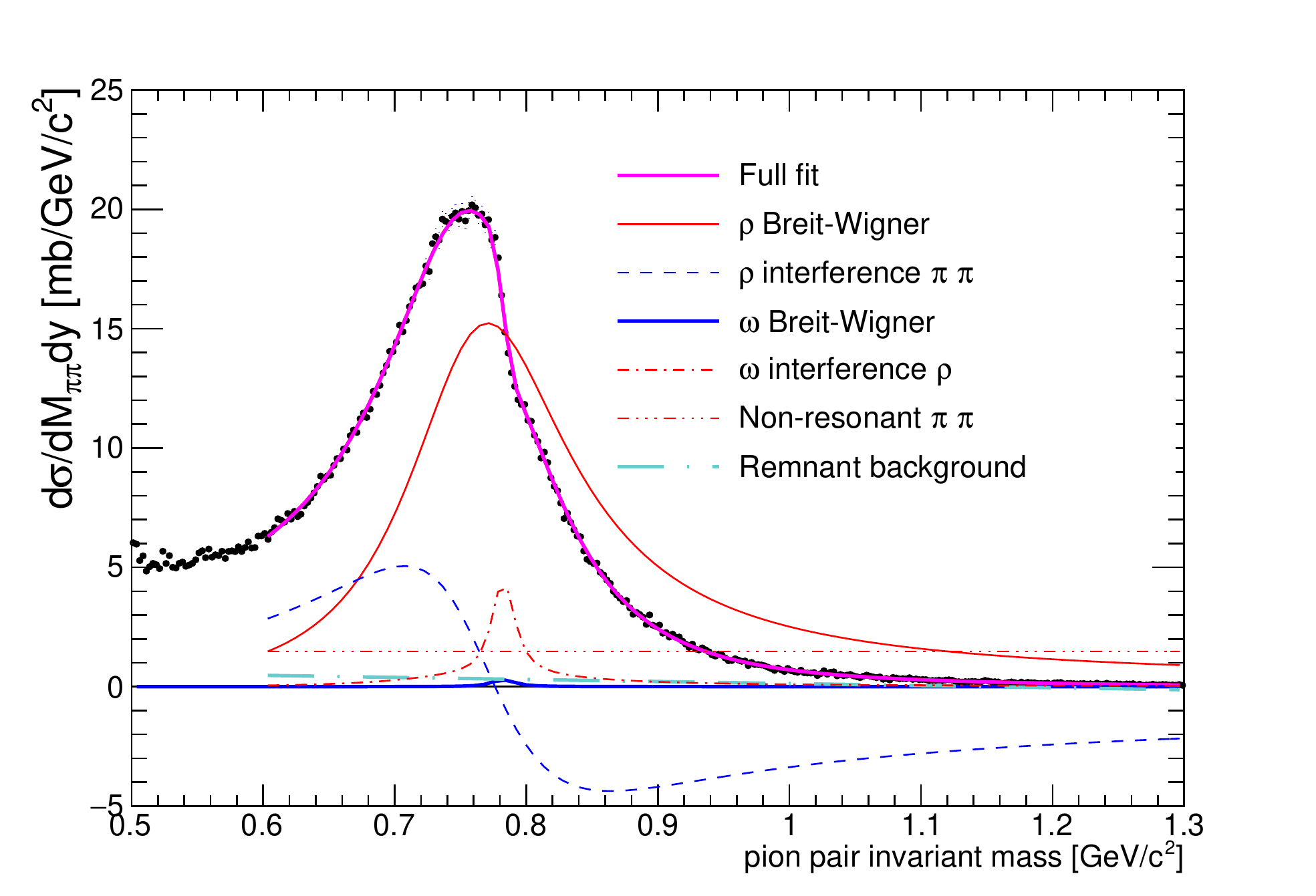} 
\includegraphics[width=0.4\columnwidth]{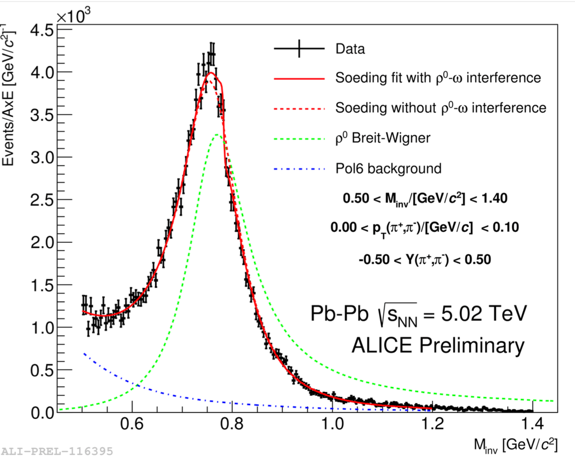}}
\label{fig:rho}
\caption{(left) The $\pi^+\pi^-$ mass spectrum seen by STAR, fit to a combination of $\rho^0\rightarrow\pi^+\pi^-$, $\omega\rightarrow\pi^+\pi^-$ and directly produced $\pi^+\pi^-$,. From  Ref. \cite{Adamczyk:2017vfu}.  (right)  The dipion mass spectrum seen by ALICE in lead-lead collisions at the LHC.  From Ref. \cite{Horak}.}
\end{figure}

Figure \ref{fig:rho} shows the mass spectrum of 294,000 exclusive dipion pairs obtained by STAR, along with a fit to a combination of $\rho^0\rightarrow\pi^+\pi^-$, $\omega\rightarrow\pi^+\pi^-$ and directly produced $\pi^+\pi^-$  \cite{Adamczyk:2017vfu}.  The two resonances were fit by Breit-Wigner functions, while the direct pions were assumed to be a continuum.  A linear function was included to model the small remnant backgrounds. In the range 600 MeV to 1.3 GeV/c$^2$, the fit had a $\chi^2/DOF$ of 255/270; without the $\omega$ term, the $\chi^2$ quadrupled.  The inclusion of the $\omega$ also improves the ALICE fit in Fig. \ref{fig:rho} (right).  The $\omega$ term itself is small, but its interference with the $\rho^0$ introduces a noticeable kink into the mass spectrum.    The  STAR ratio of direct $\pi^+\pi^-$ to $\rho^0$ is similar to multiple previous measurements and the ratio of $\omega$ to $\rho$ production and relative phase angle are similar to the previous DESY-MIT observation using 5-7 GeV photons \cite{Alvensleben:1971hz}.   The  phase angle similarity is surprising, since the DESY-MIT energy range is dominated by meson exchange, while the STAR pairs come primarily from Pomeron exchange.  STAR has also studied higher mass dipion states, finding evidence for a resonance with a mass of 1653 MeV and a width of 164 MeV.  This resonance may be consistent with the $\rho_3 (1690)$ \cite{Klein:2016dtn}.  
  
\begin{figure}[htb]
\center{\includegraphics[width=0.4\columnwidth]{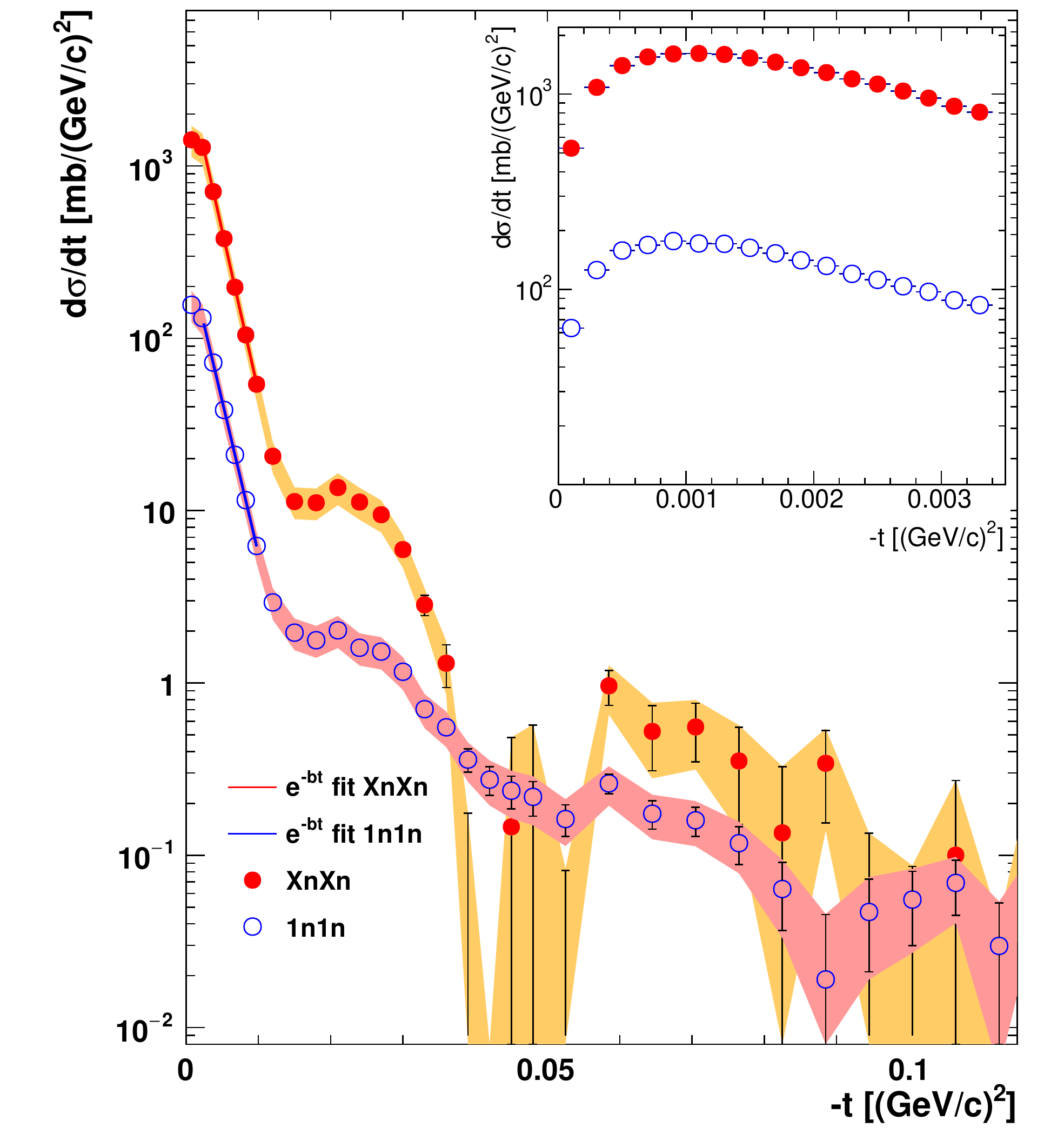} }
\label{fig:rhodsdt}
\caption{(left)  $d\sigma/dt$  for coherent $\rho^0$ photoproduction observed by STAR.  The red filled circles are for $\rho^0$ photoproduction accompanied by mutual Coulomb excitation, while the open blue circles are for events where both excitations led to single neutron emission (usually via the Giant Dipole Resonance).  The statistical uncertainties are shown by the error bars, while the shaded areas show the total uncertainty.  The inset shows a blowup of the very low $-t$ region displaying a downturn, due to destructive interference between production on the two nuclei. From  Ref. \cite{Adamczyk:2017vfu}.}
\end{figure}

The $t$-distribution from coherent $\rho$ photoproduction probes the distribution of interaction sites within the nucleus.   The $\rho^0$ is light enough so that perturbative QCD calculations are not reliable, but $\rho^0$ photoproduction is still sensitive to nuclear shadowing caused when a $q\overline q$ dipole interacts twice while traversing the target.   STAR measured $d\sigma/dt$ from coherent photoproduction by subtracting the estimated incoherent $d\sigma/dt$ from the total distribution.  The incoherent contribution was estimated by fitting the data at high $t$ (0.2 GeV$^2 < -t < 0.45 $GeV$^2$), where the coherent contribution is small, and then subtracting the fit at all $t$.   Figure \ref{fig:rhodsdt} shows the coherent $d\sigma/dt$ for two data sets, with different mutual Coulomb excitation requirements.  Two diffraction minima are visible in both data sets; the first is at $-t=0.018\pm 0.005$ GeV$^2$.  The position of this dip is a sensitive to the mean interaction radius.  Ref. \cite{Guzey:2016qwo} shows two predictions, one based directly on the nuclear form factor and the other including nuclear shadowing; the data is closer to the form factor prediction.  The inset shows a clear downturn at very small $|t|$, expected due to destructive interference between photoproduction on the two targets \cite{Klein:1999gv,Abelev:2008ew}.  STAR performed a two-dimensional (contracted in $z$) Fourier transform of $d\sigma/dt$, and explored systematic uncertainties on the distribution.  At low $b$, there was considerable variation, likely due to windowing (i.e. the exact choice of the $t-$ range to transform), and the density was negative at large $b$ because of the interference, but the positions of the edges of the density distribution were insensitive to systematic changes.

\section{Heavy quarkonium production}

Charmed and bottom quarks are heavy enough  that heavy quarkonium ($c\overline c$ or $b\overline b$) photoproduction should be describable using pQCD.  The leading-order cross-section to produce quarkonium with mass $M_v$ is \cite{Jones:2013pga}
\begin{equation}
\frac{d\sigma}{dt}\bigg|_{t=0} = \frac{\Gamma_{ee} M_V^3 \pi^3}{48\alpha} \big[\frac{\alpha_s(\overline Q^2)}{\overline Q^4} x g(x,\overline Q^2)\big]^2 \big(1+ \frac{Q^2}{M_V^2} \big)
\end{equation}
where $\overline Q^2=(Q^2+M_V^2)/4$, $x=(Q^2+M_V^2)/(W^2+Q^2)$,  $\Gamma_{ee}$ is the vector meson to photon coupling,  and $\alpha_s$ is the strong coupling constant.   The vector meson mass provides the hard scale when the photon $Q^2 \approx 0$.   

Of course, some caveats accompany this formula.  They are important in quantifying the uncertainties in using vector meson photoproduction data in fits to nuclear gluon distributions.  The reaction involves two-gluon exchange, so pQCD factorization does not strictly hold.  However, the dominant kinematic contribution occurs when the two gluons have very different $x$ values, with $x_1 \ll x_2$, so the softer gluon is not particularly important.   The two-gluon exchange can be handled exactly using a generalized parton distribution.  With the conventional gluon distribution, one can either make a smallish correction, as in Ref. \cite{Jones:2013pga}, or do the a Shuvaev transform to handle the problem exactly.    There are also corrections because the photon includes some higher order fluctuations beyond $q\overline q$.   Finally, the choice of scale, $\mu$ to evaluate the gluon distribution can significantly alter the calculated cross-section.

Another problem appears when the calculation is extended to next-to-leading order.  The leading order term includes only gluons, but the next-to-leading order terms include significant quark contributions.    The NLO `correction' is larger than the  amplitude, and has opposite sign \cite{Jones:2015nna}.  However,  there is a simple explanation for this.  In the relevant $(x,Q^2)$ region, there is no previous data.  When parton distributions are extrapolated down in $x$ to this region, they find a very small, sometimes negative gluon distribution. With additional gluons, the LO term would be larger. Using the gluon distributions determined from photoproduction data alleviates this problem.   Ref. \cite{Jones:2015nna} also investigated the scale problem.  By choosing $\mu=M_V/2$ for the LO part of the problem, the uncertainty in the cross-section due to the scale is reduced to $\pm 15\%$ - $\pm 25\%$ for $\Upsilon$ photoproduction, and somewhat more for the $J/\psi$.  Of course, the NNLO terms should be checked, but it seems that the uncertainties in the gluon distributions obtained from heavy quarkonium photoproduction data are smaller than the uncertainties in the extrapolations of fits based on other data. 

\subsection{Photoproduction on proton targets}

\begin{figure}[htb]
\center{\includegraphics[width=0.46\columnwidth]{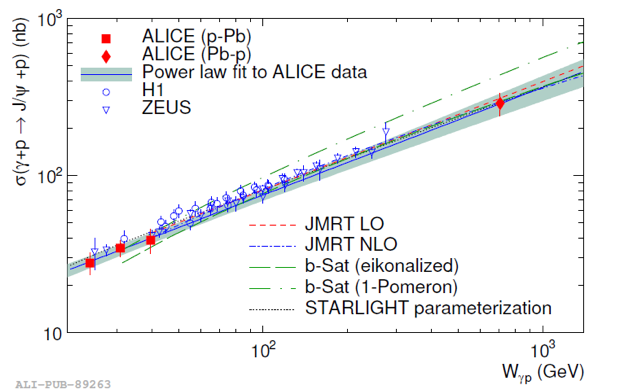} 
\includegraphics[width=0.42\columnwidth]{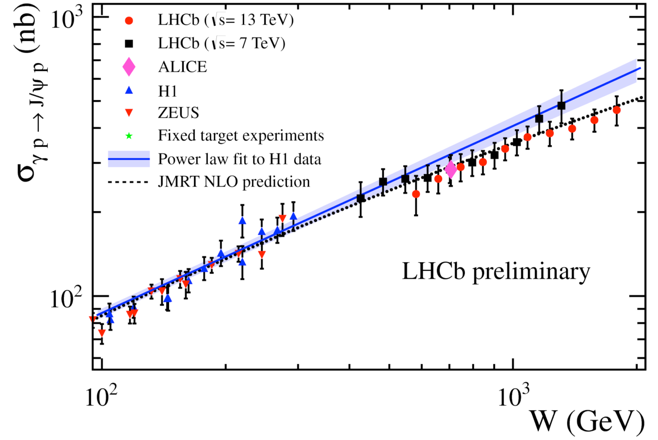} }
\caption{The $J/\psi$ photoproduction cross-sections on proton targets, from ALICE (left, from Ref. \cite{Adam:2016lmr}) and LHCb (right, from 
Ref. \cite{LHCb:2016oce}).
\label{fig:sigmajpsi}
 }
\end{figure}

ALICE has studied $J/\psi$ photoproduction in pA collisions at $\sqrt{s_{NN}}$ = 5.02 TeV \cite{Adam:2016lmr}.  Almost all of the $J/\psi$ come when photons are emitted by the gold nucleus and scatter from the proton.  The $p_t$ spectrum is dominated by the proton form factor, with only a very small low-$p_T$ peak from coherent photoproduction on gold.   Different rapidity bins, correspond to different $W_{\gamma p}$, up to $W_{\gamma p}$ = 800 GeV.   As Fig. \ref{fig:sigmajpsi} (left) shows, the data are well-fit by the STARlight parameterization \cite{Klein:2016yzr,Klein:1999qj}, which was based on HERA data. 

\begin{figure}[htb]
\center{\includegraphics[width=0.42\columnwidth]{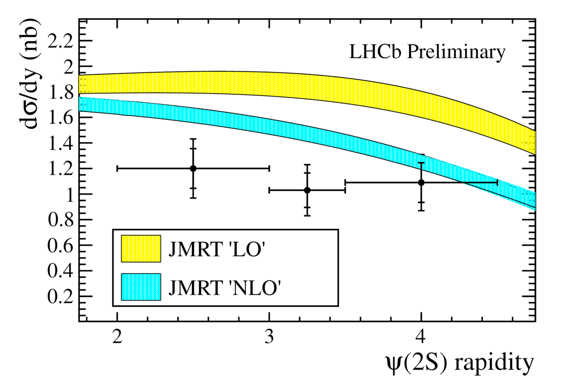} 
\includegraphics[width=0.4\columnwidth]{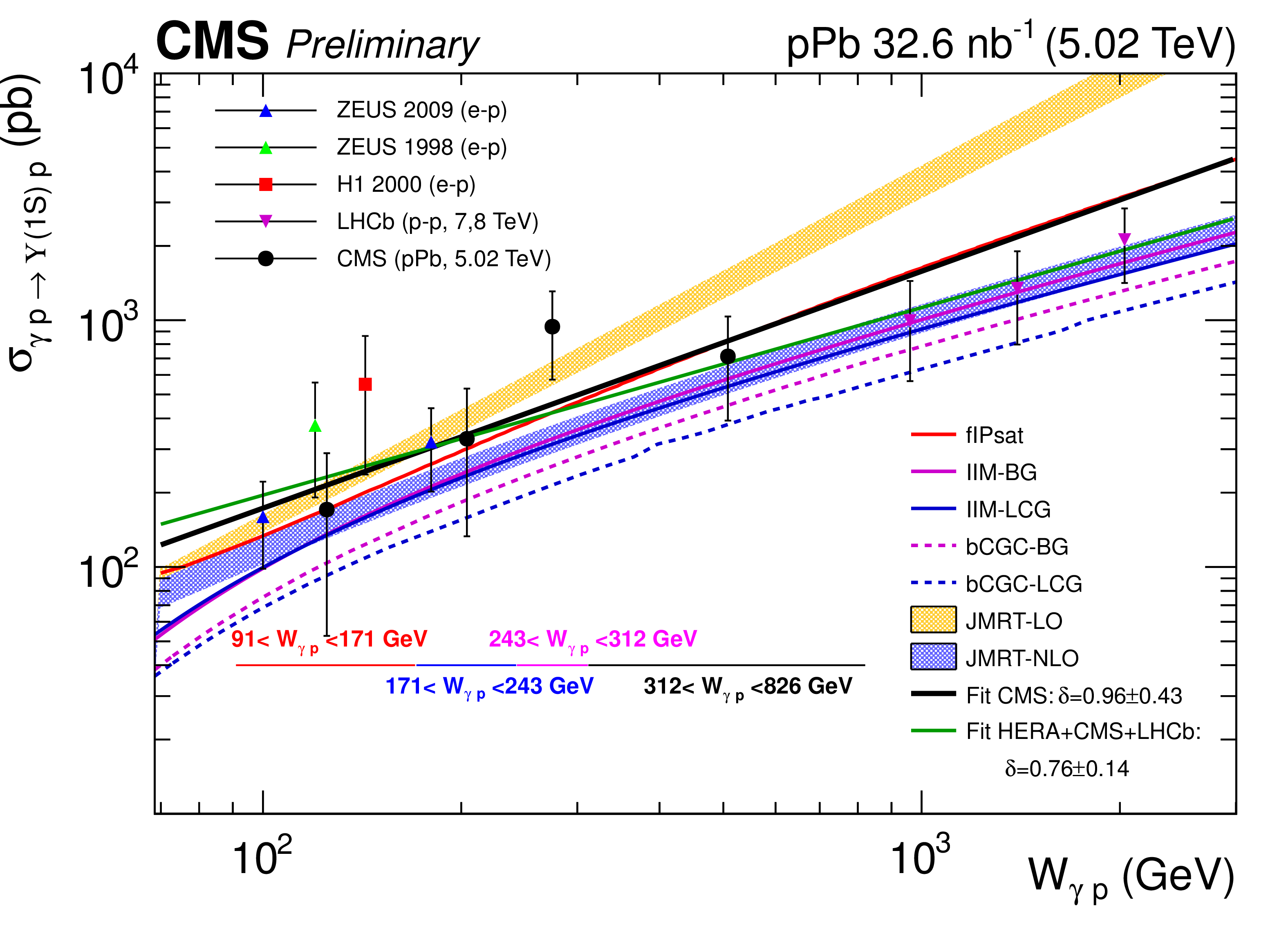} }
\caption{Photoproduction cross-sections for the $\psi'$ (left, LHCb, from Ref. \cite{LHCb:2016oce}) and $\Upsilon$(1S) (right, CMS and LHCb, from Ref. \cite{Chudasama:2016isz}).
\label{fig:sigmahq}
 }
\end{figure}

The LHCb collaboration has studied photoproduction of $\psi'$ \cite{LHCb:2016oce}, $\Upsilon (1S)$, $\Upsilon (2S)$ and $\Upsilon (3S)$ \cite{McNulty:2016sor} at far forward rapidity in $pp$ collisions, while CMS has studied $\Upsilon$ photoproduction in pA collisions \cite{Chudasama:2016isz}.   The cross-section may be written $\sigma=dN/dk_1 \sigma(k_1) + dN/dk_2 (\sigma(k_2)$ where $k_{1,2}=M_V/2 \exp(\pm y)$.  LHCb extracts the cross-section as a function of photon energy by choosing kinematic regions where $k_1$ or $k_2$ is within the HERA energy range, and then solving for cross-sections at higher or lower photon energies.    Figure \ref{fig:sigmajpsi} (right)  compares the $J/\psi$ cross-section as a function of photon energy with a power-law fit to the HERA data and with a next-to-leading order calculation.    The LHCb data taken at $\sqrt{s}=7$ TeV are in good agreement with the ALICE data and the HERA extrapolation, but the newer ($\sqrt{s}$=13 TeV) data, have slightly lower cross-sections at the highest energies. 
The CMS data is bit higher than LHCb.  The 13 TeV data are consistent with a recent NLO pQCD calculation \cite{Jones:2015nna}, but is below the HERA power-law fit.   Figure  \ref{fig:sigmahq} shows the $\psi'$ and the $\Upsilon (1S)$ data.  The same NLO calculation that describes the $J/\psi$ data also fits the $\psi'$ and the $\Upsilon (1S)$, spanning the $Q^2$ region from 2.25 GeV$^2$ to 25 GeV$^2$, a range where evidence of saturation might be expected to appear with decreasing $Q^2$. 

\subsection{$J/\psi$ photoproduction on heavy ion targets}

UPCs are the only source of high-energy heavy-target photoproduction data, and one of the few probes of low-$x$ gluon shadowing, the reduction in the gluon density in heavy nuclei (compared to protons), likely due to gluon recombination.  The $J/\psi$ is heavy enough so that its shadowing may be interpreted in terms of gluons.    Both STAR \cite{Schmidke:2016ccw} and PHENIX \cite{Afanasiev:2009hy} have studied $J/\psi$  photoproduction on gold targets at RHIC, but all of the high-statistics data are from the LHC.
ALICE has studied photoproduction on lead targets in PbPb data taken at $\sqrt{s_{NN}}$ = 2.76 and 5.02 TeV.  For the latter energy, they show a $p_T$ spectrum extending to 2.5 GeV/c, and fit it with three components: coherent photoproduction, incoherent photoproduction, and a third component where a struck nucleon dissociates.  CMS has also studied $J/\psi$ photoproduction in their PbPb data at $\sqrt{s_{NN}}$ = 2.76 TeV \cite{CMS}.   Both collaborations find cross-sections that are a bit more than half of the impulse approximation.

\begin{figure}[htb]

\center{\includegraphics[width=0.46\columnwidth]{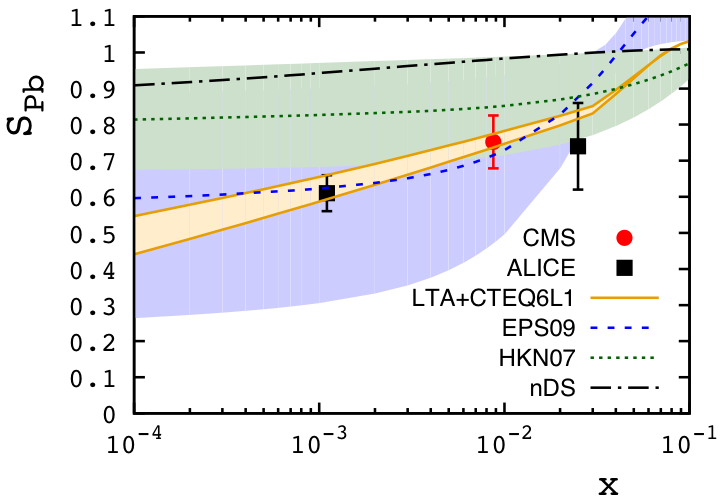}
\includegraphics[width=0.43\columnwidth]{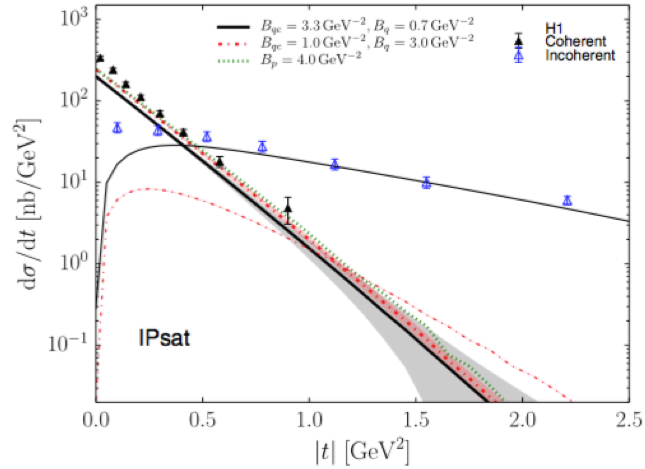}}
\caption{(left) Measurements of nuclear shadowing, extracted from the CMS and ALICE data, compared with the EPS09 parameterization.  Also shown is a leading twist calculation of shadowing.  This figure is by Vadim Guzey \cite{Frankfurt:2011cs,Guzey:2013qza}. (right) The cross-sections $d\sigma/dt$ for coherent and incoherent photoproduction of the $J/\psi$ at HERA, compared with predictions where the proton is modeled as being smooth or lumpy.  From Ref.  \cite{Mantysaari:2016jaz}.  
 \label{fig:shadowing}
 }
\end{figure}

Figure \ref{fig:shadowing} shows these cross-section data, expressed in terms of the suppression factors $S_{\rm Pb}$ related to the cross-section on proton targets, along with the predictions of the EPS09   \cite{Eskola:2009uj} and HKN07 \cite{Hirai:2007sx} shadowing parameterizations.  The data fit the EPS09 model, but somewhat below the predictions of HKN07.  The error bars on the experimental points are far smaller than the uncertainties in the EPS09 parameterization.  The newer EPPS16 parameterization has similar errors \cite{Eskola:2016oht}. These data could be used to significantly reduce the errors on these parameterizations.   The figure also shows a calculation based on leading twist shadowing \cite{Frankfurt:2011cs}, which is in good agreement with the data.  

While coherent photoproduction is sensitive to the average nuclear configuration, incoherent photoproduction probes event-by-event fluctuations, {\it i. e.} the variation in the transverse positions of the nucleons.   In the Walker-Good formalism, the total cross-section is given by \cite{Mantysaari:2016jaz}
\begin{equation}
\frac{d\sigma_{\rm tot.}}{dt} = \big< \big|A(x,Q^2,t,\Omega)\big|^2\big>_{\Omega}\
\end{equation}
where $\Omega$ represents the nuclear configurations - the individual target positions, whether the targets are nucleons, quarks or gluons.   Coherent interactions leave the nucleus unchanged, so the initial and final states must be identical, and
\begin{equation}
\frac{d\sigma_{\rm coh.}}{dt} = \big|\big< A(x,Q^2,t,\Omega)x\big>_{\Omega} \big|^2;
\end{equation}
the averaging over nuclear configurations is done for amplitudes, rather than cross-sections. The incoherent cross-section is the difference between the total cross-section and the coherent cross-section.  It is a measure of the event-by-event fluctuations in the nuclear configurations.   
A recent analysis \cite{Mantysaari:2016jaz} compared HERA data on coherent and incoherent $J/\psi$ photoproduction with calculations that included differing lumpiness or stringiness, and found that the data supported models where the proton was quite lumpy/stringy, in qualitative agreement with the large the anisotropic flow coefficients $v_2$ and $v_3$  seen in $pA$ collisions at the LHC.  

At QM2017, D. Tapai Takaki \cite{CMS} noted that when incoherent $J/\psi$ photoproduction is accompanied by neutrons, most of the neutrons are in the same hemisphere as the $J/\psi$.  This is particularly evident in the ALICE $pA$ data.   Neutrons are more likely to be produced by low-energy photons than at higher energies, consistent with a decrease in momentum transfer to the target at higher energies, and also with an approach to the 'black disk' limit, where event-by-event fluctuations slowly disappear.

\begin{figure}[htb]
\center{\includegraphics[width=0.4\columnwidth]{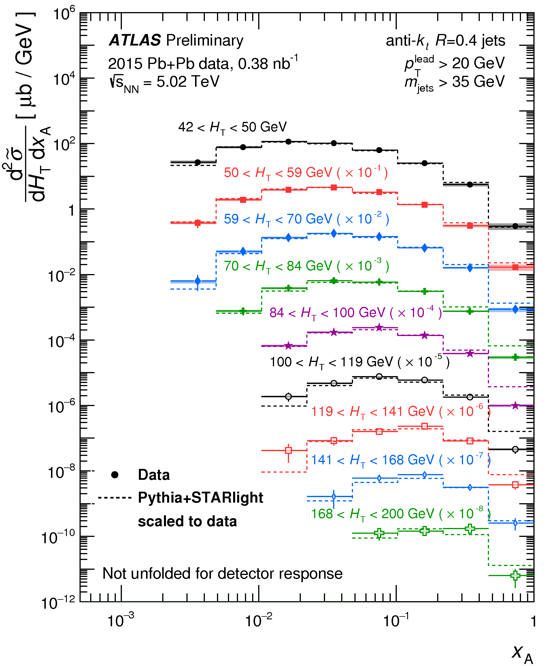} 
\includegraphics[width=0.5\columnwidth]{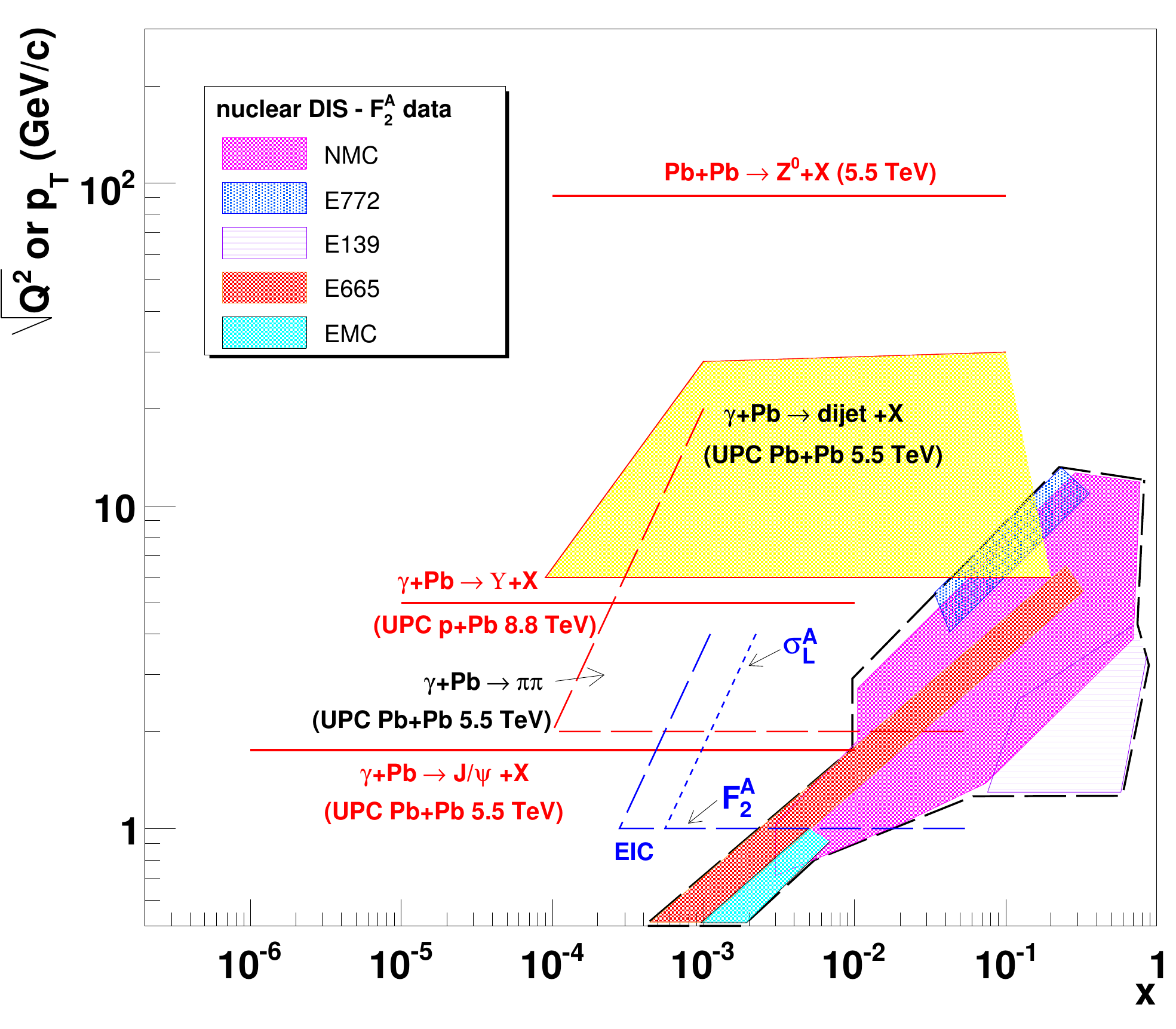}}
\caption{(left)  Dijet photoproduction data from ATLAS, compared with predictions from STARlight+PYTHIA.
From Ref. \cite{ATLAS}. (right) An overview of the ($x$,$Q^2$) regions of gluon distributions that can be probed using UPC photoproduction at the LHC. From  Ref. \cite{Baltz:2007kq}.
\label{fig:overview}
 }
\end{figure}

At QM2017, ATLAS presented initial results on the photoproduction of dijets \cite{ATLAS}.  This occurs through single gluon exchange, so it is theoretically cleaner than vector meson photoproduction.   Dijets probe a range of $x$ and $Q^2$, currently $10^{-2} < x < 1$ and $1600\ {\rm GeV}^2 < Q^2 = M_{jj}^2 < 40,000\ {\rm GeV}^2$.    Unfortunately, it is experimentally more complex, requiring the reconstruction of two jets, with only a single rapidity gap.  ATLAS selected events with one jet with $p_T > 35$ GeV, a second with $p_T>20$ GeV/c and a dijet masses in the range $42\ {\rm GeV} <M_{jj} < 200$ GeV.   They compared the data with a simulation which used STARlight \cite{Klein:2016yzr} for the photon spectrum, and PYTHIA \cite{Sjostrand:2007gs} for the jets.  ATLAS found small differences, which they are still studying.  Eventually, these data will be used to extract the gluon distribution in the nuclear target.   

\section{Conclusions and a look ahead}

Over the past few years, many exciting UPC results have appeared.  Precise measurements of two-photon production of dilepton pairs can now be used to test strong-field QED, and light-by-light photon scattering can be used to set limits on new physics processes.  Photoproduction has proven to be particularly useful as a probe of nuclear structure. The STAR Collaboration has presented a tomographic study of $\rho^0$ photoproduction which is sensitive to nuclear shadowing in gold nuclei.

Figure \ref{fig:overview} shows a 2008 projection of the region of gluon $x$ and $Q^2$ that can be studied using UPCs at the LHC \cite{Baltz:2007kq}.    The figure is remarkably prescient, accurately predicting UPC studies of heavy quarkonium and dijets.   One important advance, not considered in 2008, is the study of proton targets in $pp$ and $pA$ collisions.  Although some theoretical uncertainties remain, the $pp$ and $pA$ collisions probe gluon distributions at values an order of magnitude lower than are accessible elsewhere.  A similar picture holds for lead, with the added advantage that most of the theoretical uncertainties cancel out in calculating gluon shadowing.  Inclusion of these data in parton distribution fits will greatly reduce the uncertainties at low values of Bjorken-$x$.

Looking ahead, we anticipate more precise measurements in all of these channels, and some new ones.  Open charm photoproduction is of particular interest because is theoretically simpler than quarkonium, but can be used to probe smaller $x$ and $Q^2$ values than dijet photoproduction.   We also anticipate that the RHIC $\rho^0$ tomographic approach will be extended to include $J/\psi$s at the LHC.  The $J/\psi$ is heavy enough to probe gluons directly, so these tomography studies can directly image the gluon distributions in the nucleus.  At RHIC, we look forward to the advent of photoproduction studies on polarized proton targets.  One particular attraction is the first measurement of the generalized parton distribution E (GPD-E), which can be measured from the left-right asymmetries in $J/\psi$ photoproduction on polarized protons \cite{Schmidke:2016ccw}.

We also anticipate more connections with hadronic collisions.  ALICE and STAR have observed an excess of dilepton pairs in peripheral collisions (40\% to 80\% centrality for STAR, 70\% to 90\% centrality for ALICE), concentrated at $p_T< 100$ MeV/c, at rates well above those expected for hadronic processes, so it seems fairly safe to attribute this excess to photoproduction.  The photoproduction data could provide additional information about the event, including on the direction of the event plane. 

I thank Vadim Guzey for making Fig. 6 (left) for this presentation.  This work was funded by the U.S. Department of Energy under contract number DE-AC-76SF00098. 




\end{document}